\documentclass{ragtime}
\usepackage{bm}
\title[Application of a symplectic integrator in a relativistic system]%
      {Application of a symplectic integrator in a non-integrable relativistic system}
\author[O. Kop\'{a}\v{c}ek, 
        V. Karas,
	  J. Kov\'a\v{r}
        and Z. Stuchl\'{i}k]
       {Ond\v{r}ej Kop\'{a}\v{c}ek\at{1,2,a}
Vladim\'{i}r Karas \at{1}
Ji\v{r}\'{\i} Kov\'a\v{r}\at[]{3}\splitauthors 
and Zden\v{e}k Stuchl\'{\i}k\at[]{3}\\
        \ins{1}Astronomical Institute of the Academy of Sciences of the Czech Republic,\splitins[1]
                                                 Bo\v{c}n\'{i} II 1401/1a, CZ-141\,31 Prague,
        Czech Republic\\
        \ins{2}Faculty of Mathematics and Physics of Charles University,\splitins[1] Ke Karlovu 3, CZ-121\,16 Prague,
        Czech Republic\\
\ins{3}Institute of Physics, Faculty of Philosophy \& Science,
        Silesian University in Opava,\splitins[1]
        Bezru\v{c}ovo n\'am.~13, CZ-746\,01 Opava,
        Czech Republic\\
\ins{a}\Email{kopacek@ig.cas.cz}}

\newcommand{\rff}[1]{Fig.\,\ref{#1}}

\begin{document}
\begin{abstract}
  We present a detailed comparison of several integration schemes applied to the dynamic system consisting of a charged particle on the Kerr background endowed with the axisymmetric electromagnetic test field. In particular, we compare the performance of the symplectic integrator with several non-symplectic routines and discuss under which circumstances we should choose the symplectic one and when we should switch to some other scheme. We are basically concerned with two crucial, yet opposing aspects -- accuracy of the integration and CPU time consumption. The latter is generally less critical in our application while the highest possible accuracy is strongly demanded.  
\end{abstract} 

\section{Introduction}\label{intro}
In our recent study of the test particle dynamics \citep{kopacek10,kovar10} we faced the problem of numerical integration of relativistic dynamic system described by the non-integrable equations of motion. Such system generally allows for both regular and chaotic orbits. We first applied several standard 'all-purpose' integration routines to realize that they are unable to provide sufficiently accurate results concerning the long-term integration. Seeking for the scheme which would better fit our problem and provide more reliable results we finally employed symplectic integrators which are specifically designed for the integration of Hamiltonian systems. 

In this contribution we compare performance of a symplectic routine with several non-symplectic integrators. We treat separately the case of regular and chaotic motion because we may expect different results. Particular system which we employ in the survey consists of a charged test particle orbiting above the outer horizon of the Kerr black hole which is immersed into the asymptotically uniform magnetic field aligned with the rotation axis \citep{wald74}. Specification of this system along with the detailed study of the charged particle dynamics is given by \citet{kopacek10}. Current paper is based on the results previously published in the Ph.D. thesis of one of the authors \citep{thesis}.

We recall that in the given system the particle of rest mass $m$ is characterized by its specific angular momentum $\tilde{L}\equiv{}L/m$, specific energy $\tilde{E}\equiv{}E/m$ and specific charge $\tilde{q}\equiv{}q/m$. Black hole of mass $M$ is described by the spin parameter $a$ and specific test charge $\tilde{Q}\equiv{}Q/M$. Background magnetic field is specified by its asymptotic strength $B_{0}$. Inspecting the equations of motion we reveal that $\tilde{q}$, $\tilde{Q}$ and $B_{0}$ are not independent variables and we only need to specify values of products $\tilde{q}\tilde{Q}$ and $\tilde{q}B_{0}$ to characterize the system. We use standard Boyer-Lindquist coordinates $x^{\mu}= (t,\:r, \:\theta,\:\varphi)$ and denote the canonical four-momentum as $\pi_{\mu}=(\pi_t,\pi_r,\pi_{\theta},\pi_{\varphi})$. Standard kinematical four-momentum $p^{\mu}$ and canonical four-momentum are related as follows $p^{\mu}=\pi^{\mu}-qA^{\mu}$ where $A^{\mu}$ stands for the electromagnetic four-potential. Integration variable is affine parameter $\lambda$ defined as $\lambda\equiv\tau/m$ where $\tau$ denotes the
proper time of the particle. We use geometrized units $G=c=1$ and scale all quantities by the mass of the black hole $M$.

We are dealing with the integration of the autonomous Hamiltonian system$^1$ \footnotetext[1]{Equations of motion may be equivalently expressed in terms of Lorentz force \citep[][,~p.~898]{mtw} which leads to the set of four second order ODEs. Numerical experiments, however, led us to the conclusion that this formulation is computationally less effective compared to the Hamiltonian formalism. Generally for a given numerical scheme with the same parameters (resulting in similar accuracy of the integration) the integration of Hamilton's equations was roughly two times faster. Moreover, the symplectic methods may only be applied in the Hamiltonian formulation of the problem.} whose equations of motion form a specific subclass of first order ordinary differential equations (ODEs). Two fundamental characteristics of the Hamiltonian flow should be highlighted 
 \begin{itemize}
\item conservation of the net energy (Hamiltonian) of the system
\item conservation of the symplectic 2-form $\bm{\omega}=\mathbf{d}\pi_{\mu}\wedge \mathbf{d}x^{\mu}$.
\end{itemize}
Here $\mathbf{d}$ stands for the exterior derivative and $\wedge$ denotes the wedge product.

In the classical mechanics the natural choice of the generalized coordinates leads to the Hamiltonian which may be interpreted as a net energy of the system. This is true even for the system of a charged particle in the external electromagnetic field where the generalized momenta-dependent potential is introduced \citep[][,~chap.~ 8]{goldstein}. Time-independance of the Hamiltonian is thus equivalent to the conservation of the net energy of the system. In the general relativistic version of this system, however, we employ super-hamiltonian formalism \citep[][,~chap.~21]{mtw} in which the energy of the particle $E$, as a negatively taken time component of the canonical momentum $E\equiv -\pi_t$, is conserved by virtue of the Hamilton's equations itselves providing that the super-hamiltonian doesn't depend on the coordinate time $t$. On the other hand the value of the super-hamiltonian $\mathcal{H}=\frac{1}{2}p_{\mu}p^{\mu}$ is by construction equal to $-\frac{1}{2}m^2$ where $m$ is the rest mass of the particle. Conservation of the super-hamiltonian in the system is thus equivalent to the conservation of the rest mass of the particle.

\begin{figure}[htb]
\centering
\includegraphics[scale=0.54,clip]{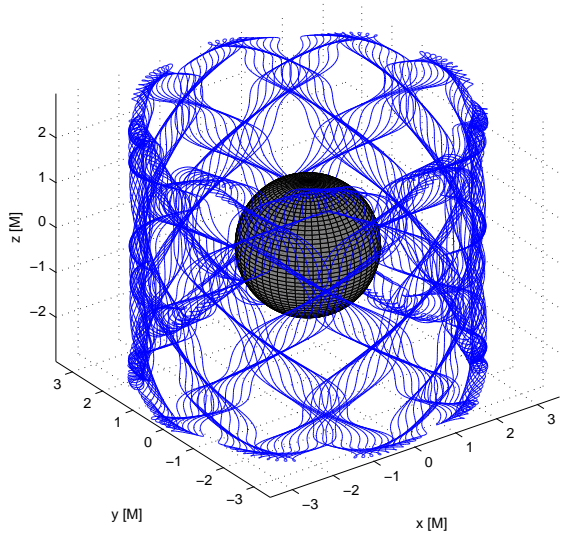}
\caption{Regular trajectory of a charged test particle ($\tilde{q}\tilde{Q}=1$, $\tilde{L}=6\;M$ and $\tilde{E}=1.6$) on the Kerr background
($a=0.9\;M$) with Wald magnetic field
($\tilde{q}B_{0}=1M^{-1}$). The particle is
launched at $r(0)=3.68\:M$, $\theta(0)=1.18$ with $u^r(0)=0$.}
\label{traj_reg_3d}
\end{figure}

By conservation of the symplectic 2-form $\bm{\omega}$ we mean that its components $\omega_{\alpha\beta}$ in the basis $\left(\mathbf{d}t(\lambda),\mathbf{d}r(\lambda),\mathbf{d}\theta(\lambda),\mathbf{d}\varphi(\lambda),\mathbf{d}\pi_t (\lambda),\mathbf{d}\pi_r (\lambda), \mathbf{d}\pi_{\theta} (\lambda), \mathbf{d}\pi_{\varphi}(\lambda) \right)$ do not change during the evolution of the system and for arbitrary value of the affine parameter $\lambda$ (i.e. at each point of the phase space trajectory) we obtain
\begin{equation}
\omega_{\alpha\beta}=
\begin{pmatrix}
0&-\mathbb{I}\\
\mathbb{I}&0\\
\end{pmatrix},
 \label{symplmatice}
\end{equation}
where $\mathbb{I}$ stands for the four-dimensional identity submatrix and $0$ is the null submatrix of the same dimension. Conservation of the symplectic structure expresses in the abstract geometrical language the fact that the evolution of the system is governed by the Hamilton's canonical equations. See \citet{arnold} for details on the geometric formulation of the Hamiltonian dynamics. 

It would be highly desirable to use such integration scheme which would conserve both quantities which are conserved by the original system. It appears, however, that this is not possible for non-integrable systems and one has to decide whether he employs the scheme which conserves energy or rather the integrator which keeps the symplectic structure. The latter are referred to as symplectic integrators and by many accounts provide most reliable results in numerical studies involving Hamiltonian systems. See \citet{yoshida93} for a comprehensive review on symplectic methods.

\begin{figure}[hp]
\centering
\includegraphics[scale=0.56,clip]{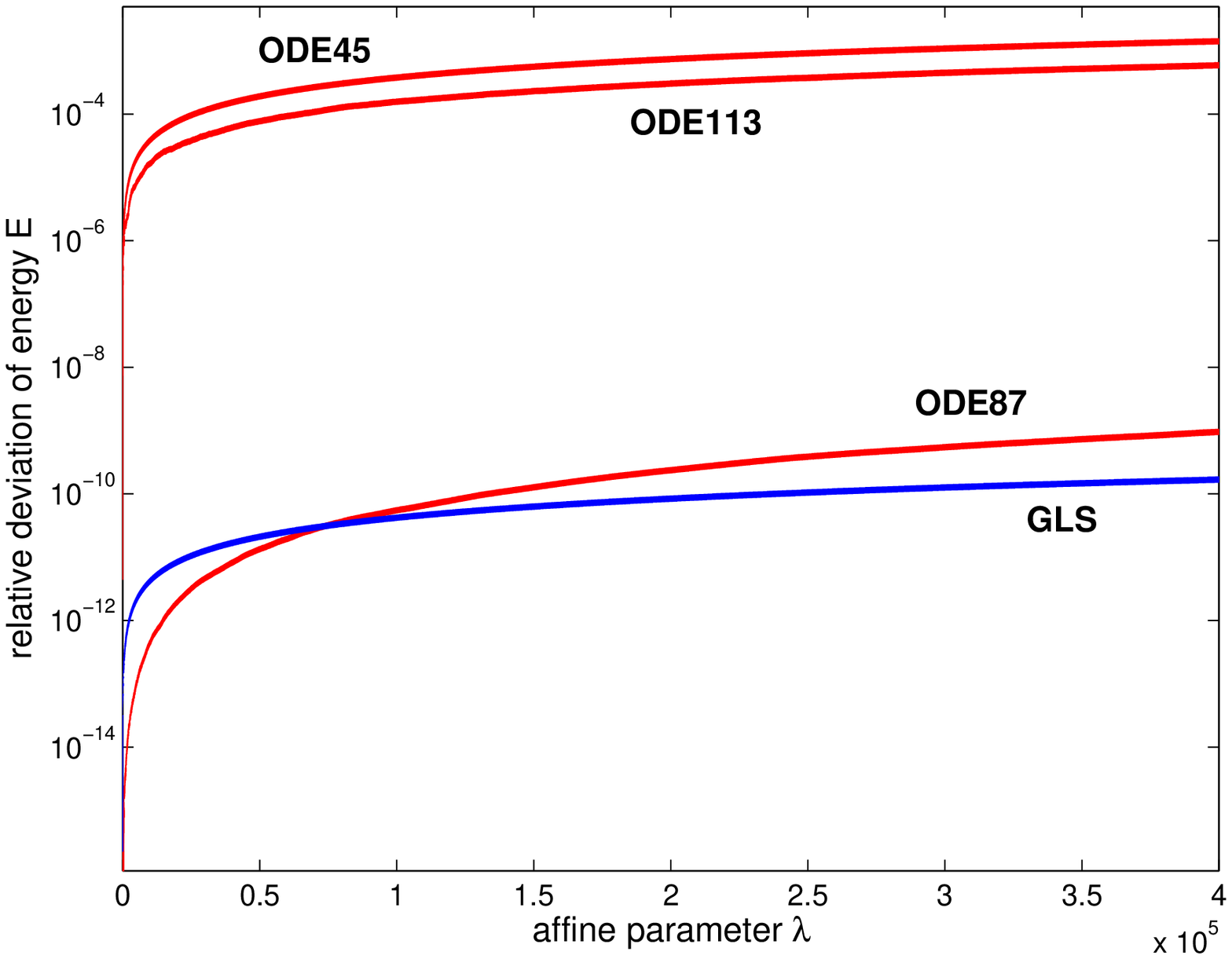}
\includegraphics[scale=0.56,clip]{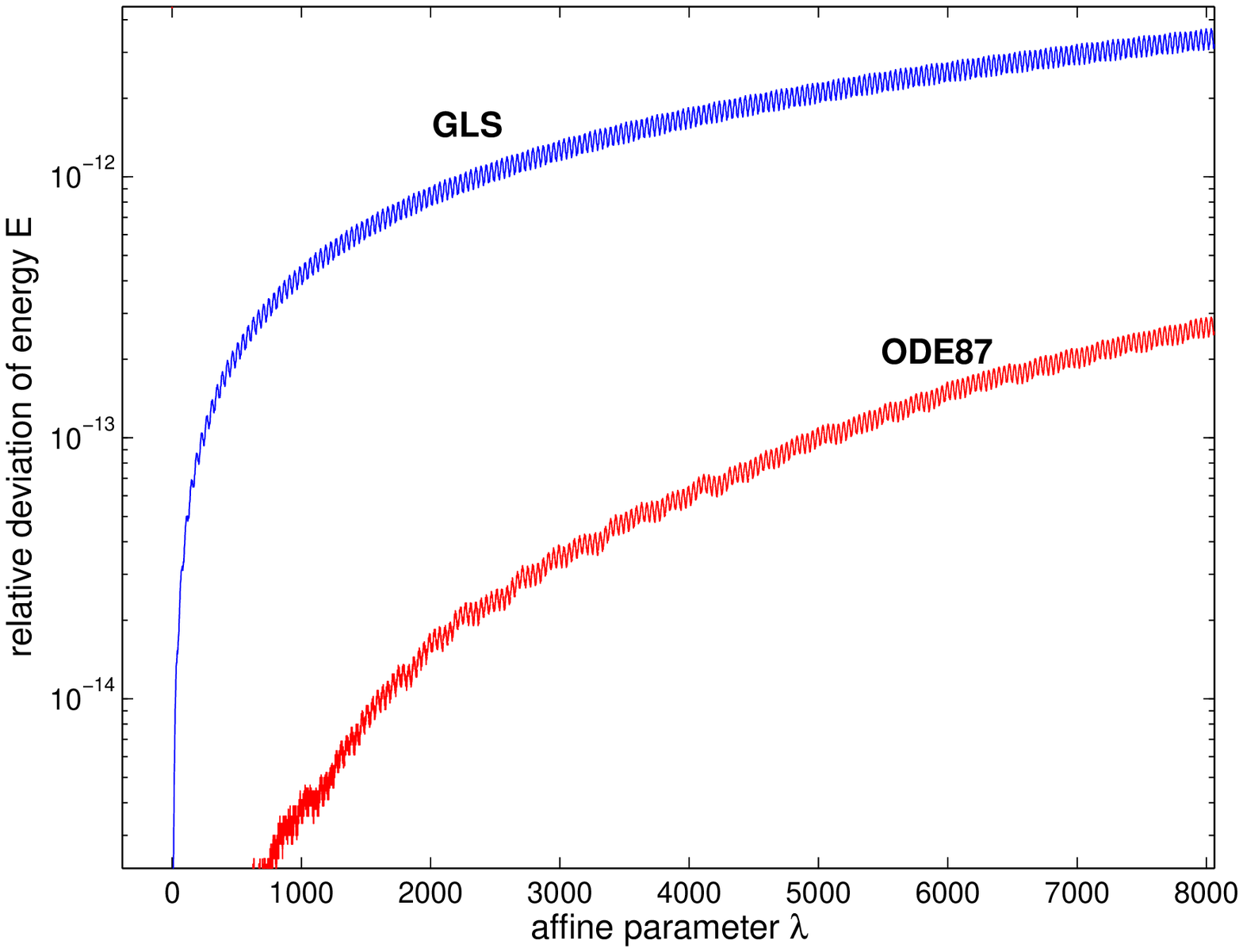}
\caption{Comparison of the integrators in the case of regular trajectory. Symplectic GLS provides the most reliable results for $\lambda\gtrsim10^5$. Bottom panel shows that besides secular drift in energy (artificial excitation or dumping of the system; plot shows absolute values, however) it also oscillates on the short time scale.}
\label{traj_reg}
\end{figure}

We list all the schemes we employ in this survey specifying their basic properties. We shall actually compare one symplectic method with several standard integrators. Code names we use for the schemes are those which denote the routines in the MATLAB system.
\begin{itemize}
\item
GLS -- Gauss-Legendre symplectic solver, $s$-stage implicit Runge-Kutta (RK) method, crucial control parameter: stepsize $h$
\item
ODE87 -- Dormand-Prince 8th - 7th order explicit RK scheme, the most precise RK method (local error of order $O(h^8)$), adaptive stepsize -- RelTol is set to control the local truncation error  
\item
ODE113 -- multistep Adams-Bashforth-Moulton solver, based on the pre\-dic\-tor-corrector method (PECE), RelTol is set
\item
ODE45 -- Dormand-Prince seven stage 5th-4th order method of explicit RK family, adaptive stepsize, default integration method in MATLAB and GNU OCTAVE, error is controlled by RelTol
\end{itemize}
Apart from ODE113 all other routines are single-step (Runge-Kutta like) methods which means that they express the value of the solution in the next step in terms of a single preceding step. They may be related explicitly or implicitly. Multistep methods in contrast employ more preceding steps to calculate the solution at the succeeding point. RelTol (relative tolerance) is a parameter which specifies the highest allowed relative error in each step of integration (local truncation error) when the adaptive stepsize methods are used. In the case of exceeding the RelTol the stepsize is reduced automatically to decrease the error.

We comment that for general non-separable Hamiltonians only implicit symplectic schemes may be found. Explicit methods exist for separable Hamiltonians and for some special forms of non-separable ones \citep{chin09}. Besides other implications of the usage of the implicit methods we note that they necessarily involve some type of iterative scheme which is typically of a Newton's type and thus requires to supply Jacobian of the right hand sides of the equations of motion which is the Hessian matrix of the second derivatives of the super-hamiltonian $\mathcal{H}$ in our case.

\begin{table}[htb]
\label{stat_reg}
\begin{center}
\begin{tabular}{l|l|l|l|l}
integrator& $\Delta |E|/|E|$ & $t_{\rm{comp}} [h]$& RelTol& stepsize $h$\\
\hline
GLS&$\approx 10^{-10}$ & 14 & N/A & 0.25\\
ODE87&$\approx 10^{-9}$& 14 & $10^{-14}$ & adaptive\\
ODE113&$\approx 10^{-3}$& 1/3  & $10^{-14}$ & adaptive\\
ODE113&$\approx 10^{-3}$& 1/4 & $10^{-6}$ & adaptive\\
ODE45&$\approx 10^{-3}$& 1/4 & $10^{-14}$ & adaptive\\
\end{tabular}
\caption{Comparison of the performance of several integration schemes for the regular trajectory integrated up to $\lambda=4\times 10^{5}$ (see \rff{traj_reg}). Quantity $t_{\rm{comp}}$ expresses the CPU time in hours.}
\end{center}
\label{tab_reg}
\end{table}

\begin{figure}[htb]
\centering
\includegraphics[scale=0.54, clip]{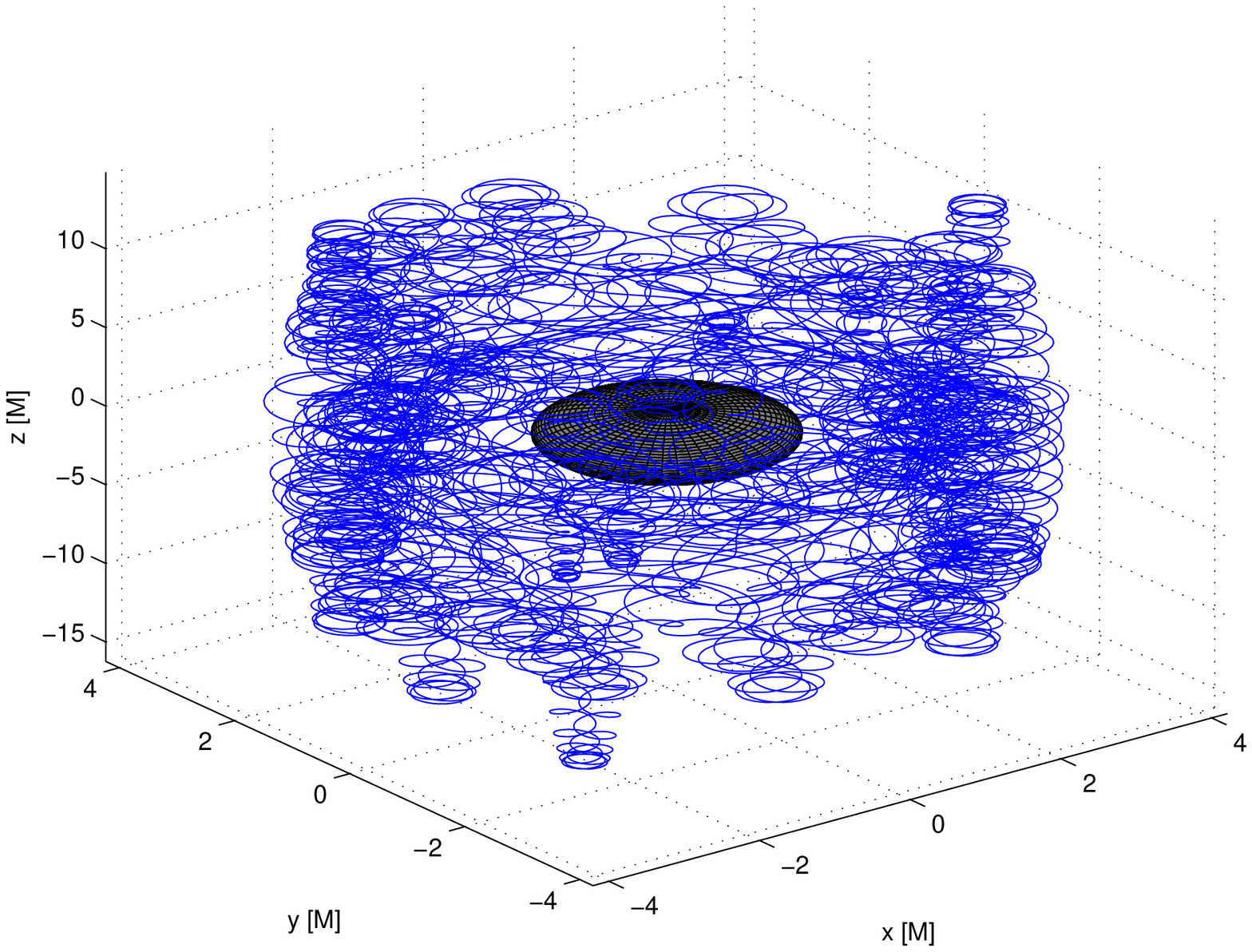}
\caption{Chaotic trajectory of a charged test particle ($\tilde{q}\tilde{Q}=1$, $\tilde{L}=6\;M$ and $\tilde{E}=1.8$) on the Kerr background
($a=0.9\;M$) with Wald magnetic field
($\tilde{q}B_{0}=1M^{-1}$). The particle is
launched at $r(0)=3.68\: M$, $\theta(0)=1.18$ with $u^r(0)=0$.}
\label{traj_chaos_3d}
\end{figure}

Another inconvenience connected with the symplectic methods is their failure to conserve the symplectic structure once the adaptive stepsize method would be used \citep{skeel92}. Therefore the stepsize has to be set rigidly for a given integration segment when using symplectic method. Several workarounds have been suggested to combine benefits of symplectic solvers and variable stepsize algorithms -- e.g. Hairer's symplectic meta-algorithm \citep{hairer97} which is, however, only applicable to the separable Hamiltonians. In our context one would considerably suffer from the fixed timestep only in the case of highly eccentric orbits.

\begin{figure}[hp]
\centering
\includegraphics[scale=0.56,clip]{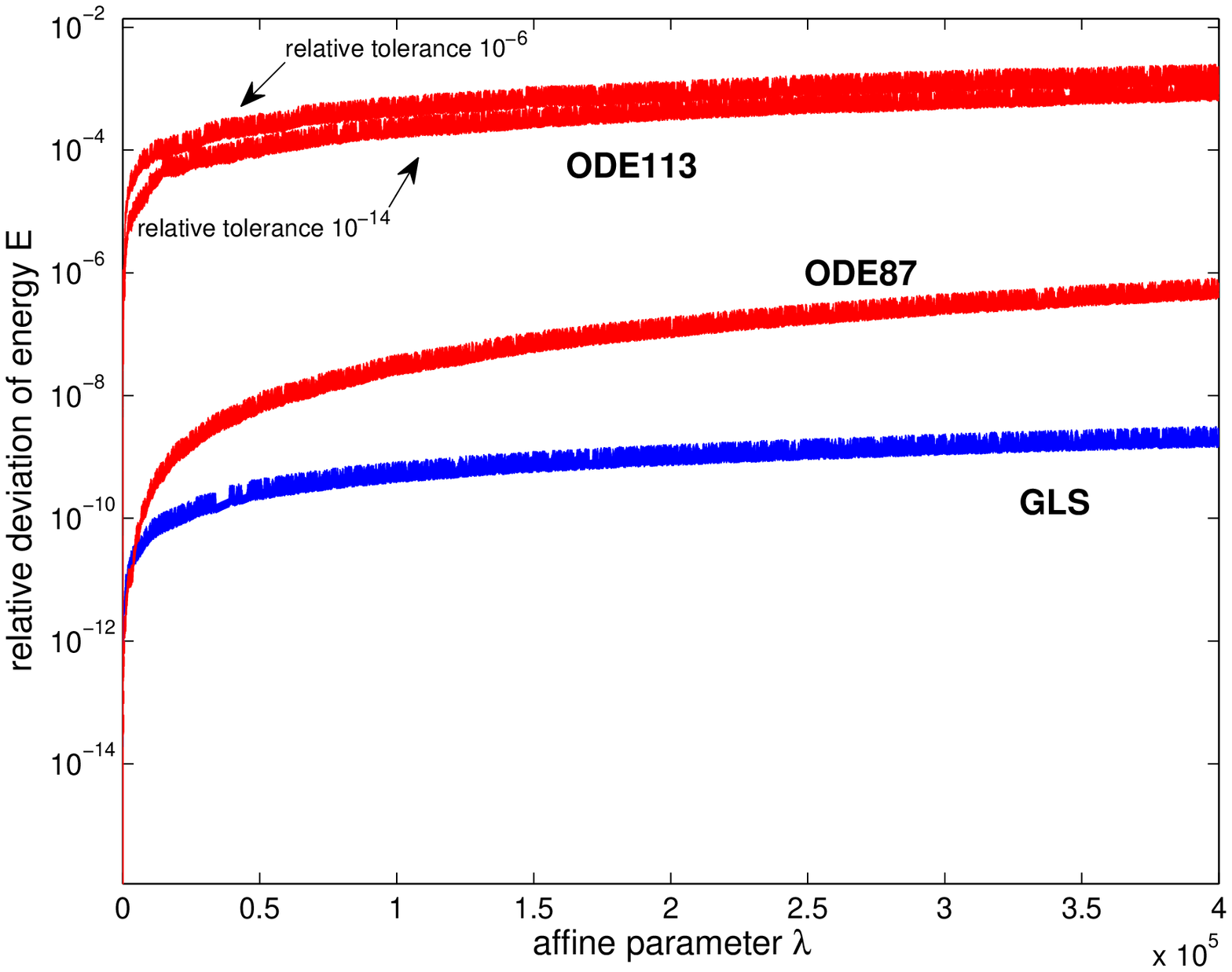}
\includegraphics[scale=0.56,clip]{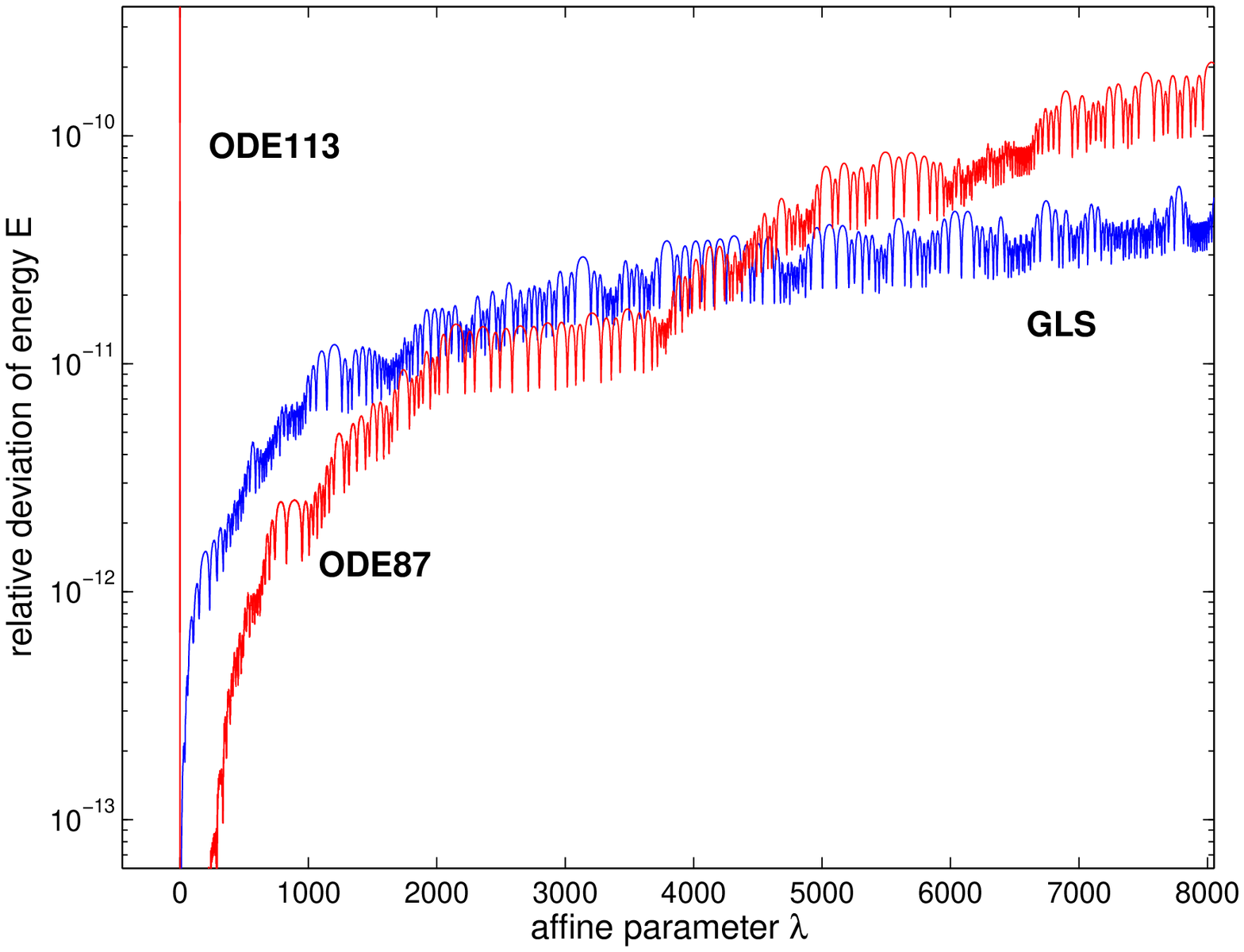}
\caption{Comparison of the integrators in the case of chaotic trajectory. For $\lambda\gtrsim5\times10^3$ the GLS dominates in accuracy over other schemes with the difference rising steadily. In the upper panel we compare ODE113's outcome for two distinct values of the RelTol parameter. ODE45 is not shown to avoid overlapping of its plot with ODE113 curves.}
\label{traj_chaos}
\end{figure}

\section{Performance of the integrators}\label{comparison}

First we integrate the cross-equatorial regular trajectory depicted in \rff{traj_reg_3d}. Comparison of the performance of the integrators is plotted in \rff{traj_reg}. We plot relative deviation of the particle's specific energy $\tilde{E}$ from its initial value rather than the error in super-hamiltonian because the discussion of motion in \citet{kopacek10} was mostly held in terms of $\tilde{E}$ whose impact upon the trajectory is thus more familiar to us. We calculate the current value of $\tilde{E}$ from the super-hamiltonian $\mathcal{H}$, while the value of $\pi_t$ remains truly constant regardless the integrator since the Hamilton's equation for its evolution is simply $\mathrm{d}\pi_t / \mathrm{d}\lambda =0$.

Stepsize of GLS is set in such a way that the integration consumes roughly the same amount of the CPU time as it does for ODE87 with $\rm{RelTol}=10^{-14}$ to make the results comparable. The global accuracy of the GLS solver could be further increased by reducing the stepsize while decreasing the RelTol hardly improves the secular accuracy of non-symplectic methods here (we have compared $\rm{RelTol}=10^{-6}$ and $\rm{RelTol}=10^{-14}$ results for ODE113 obtaining global errors of the same orders in both cases). 

\begin{table}[htb]
\label{stat_chaos}
\begin{center}
\begin{tabular}{l|l|l|l|l}
integrator& $|\Delta{}E|/|E|$ & $t_{\rm{comp}}\; [h]$& RelTol& stepsize $h$\\
\hline
GLS&$\approx 10^{-9}$ & 14 & N/A & 0.25\\
ODE87&$\approx 10^{-6}$& 14 & $10^{-14}$ & adaptive\\
ODE113&$\approx 10^{-3}$& 1/6  & $10^{-14}$ & adaptive\\
ODE113&$\approx 10^{-3}$& 1/6 & $10^{-6}$ & adaptive\\
ODE45&$\approx 10^{-3}$& 1/2 & $10^{-14}$ & adaptive
\end{tabular}
\caption{Comparison of the performance of several integration schemes for the chaotic trajectory integrated up to $\lambda=4\times 10^{5}$ (see \rff{traj_chaos}).}
\end{center}
\label{tab_chaos}
\end{table} 

We observe that the error of GLS rises steeply at the beginning and ODE87 is considerably better for some amount of time. However then the error of GLS almost saturates while ODE87's error keeps growing significantly. For $\lambda\gtrsim10^5$ which corresponds to $\approx 1000$ revolutions around the center$^2$ \footnotetext[2]{For instance for $M=10^{6}M_{\odot}$ the azimuthal proper period of a given particle reads $T_{\varphi} \approx 10^3 s$ in SI.} the GLS scheme becomes more accurate than ODE87 with the difference further rising steadily. We conclude that in the case of regular trajectory ODE87 is appropriate for short-term accurate integration and GLS for any longer accurate integrations. On the other hand for fast, though inaccurate computations one employs ODE113 on all time scales. See Table \ref{tab_reg} for the summary.

\begin{figure}[htb]
\centering
\includegraphics[scale=0.35,clip]{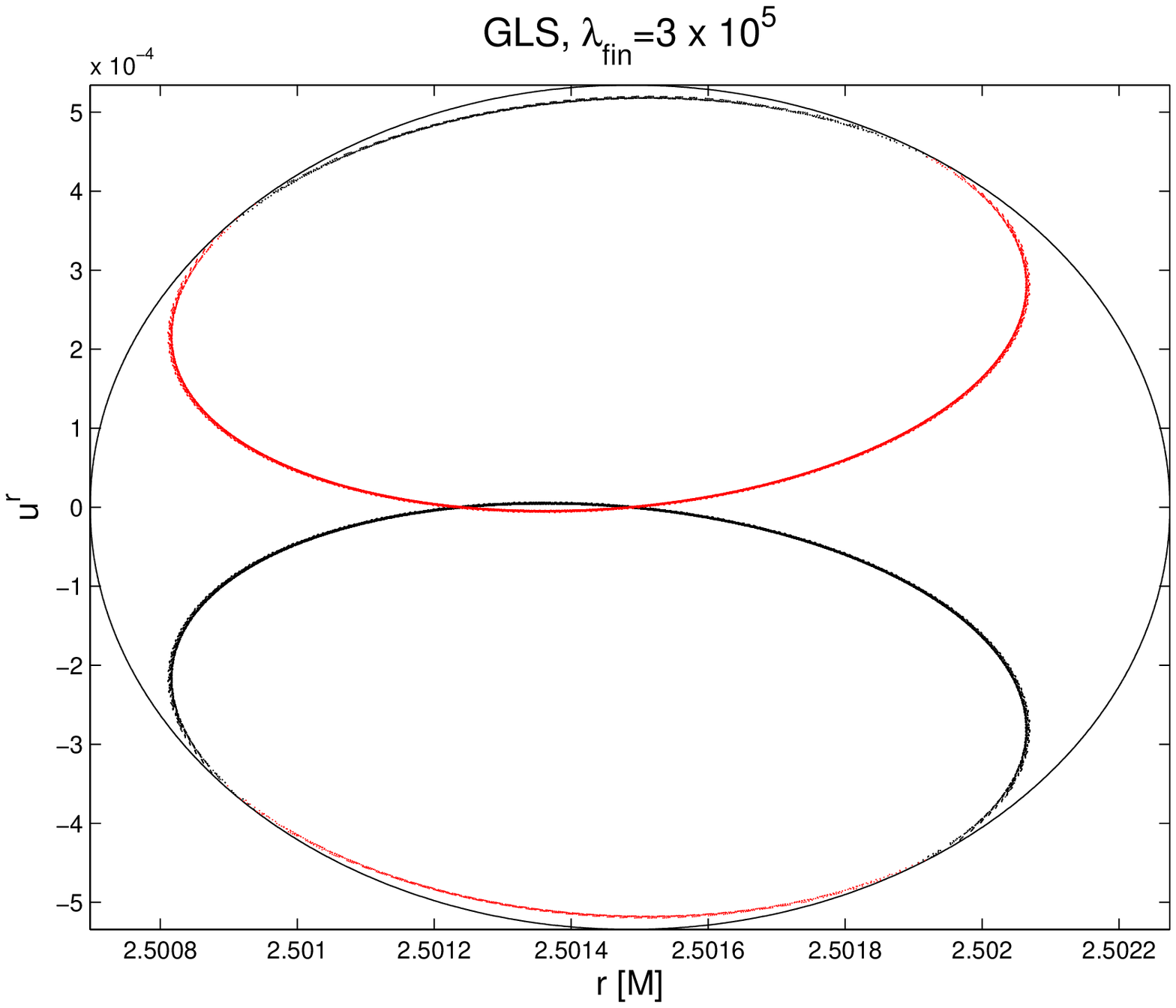}
\includegraphics[scale=0.35,clip]{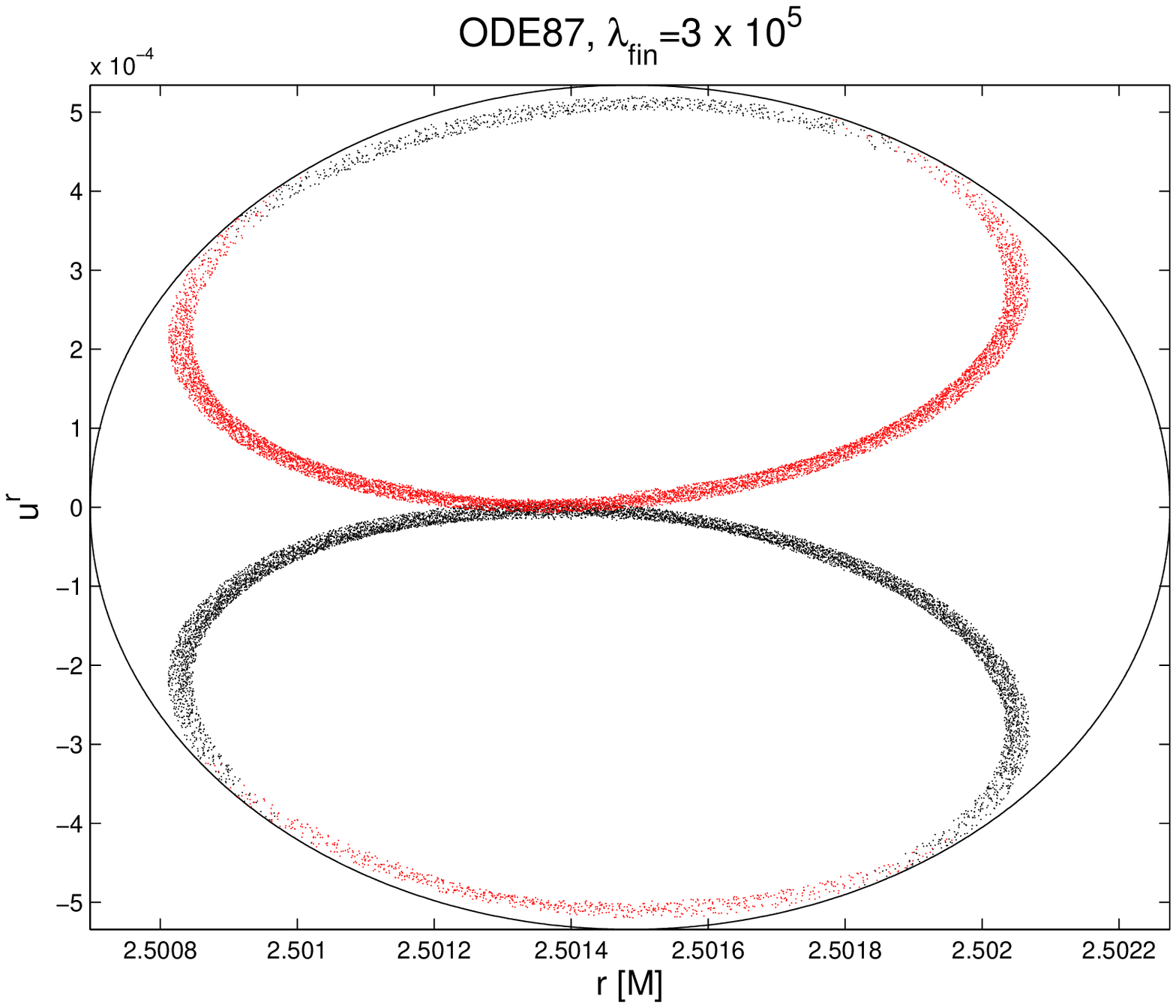}
\includegraphics[scale=0.35,clip]{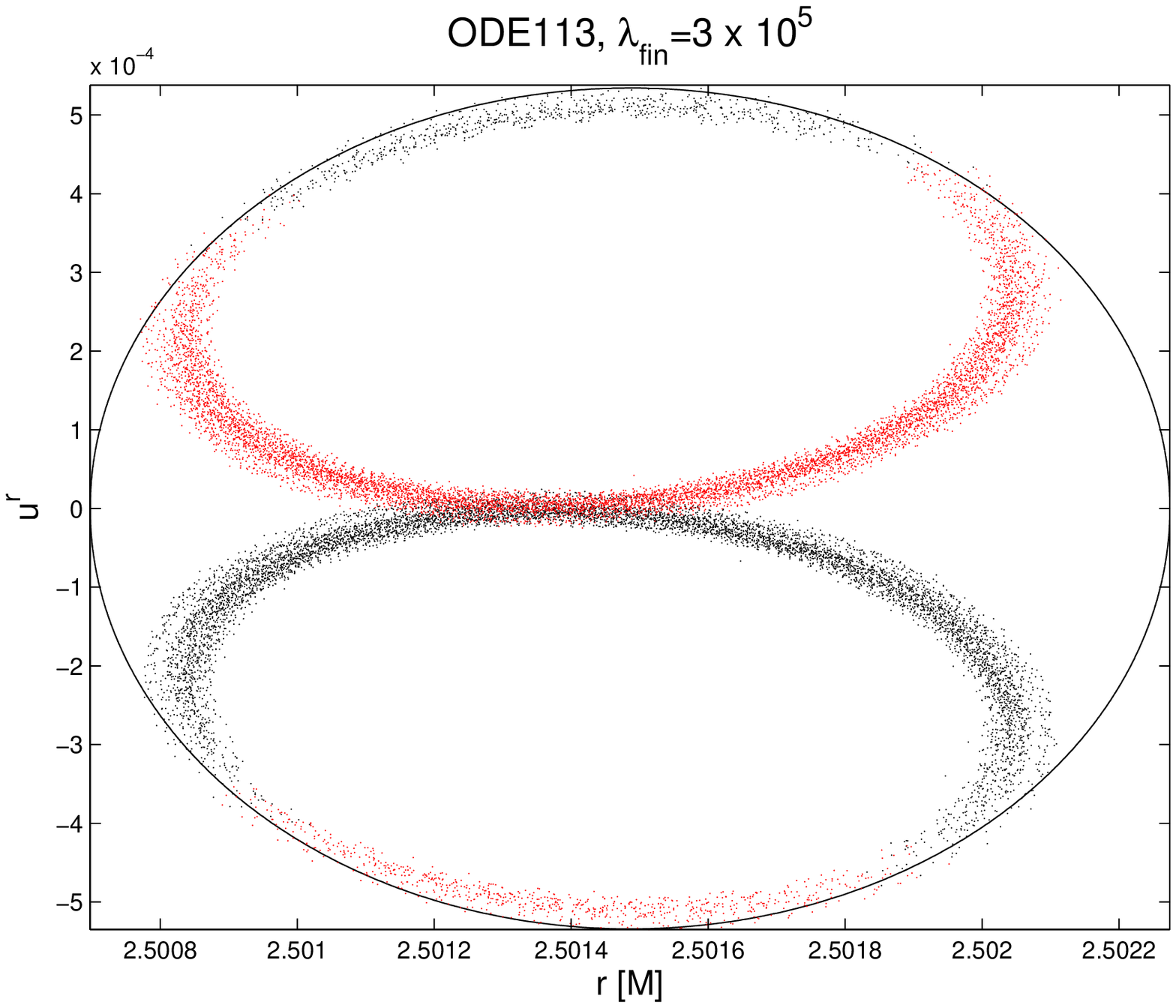}
\includegraphics[scale=0.35,clip]{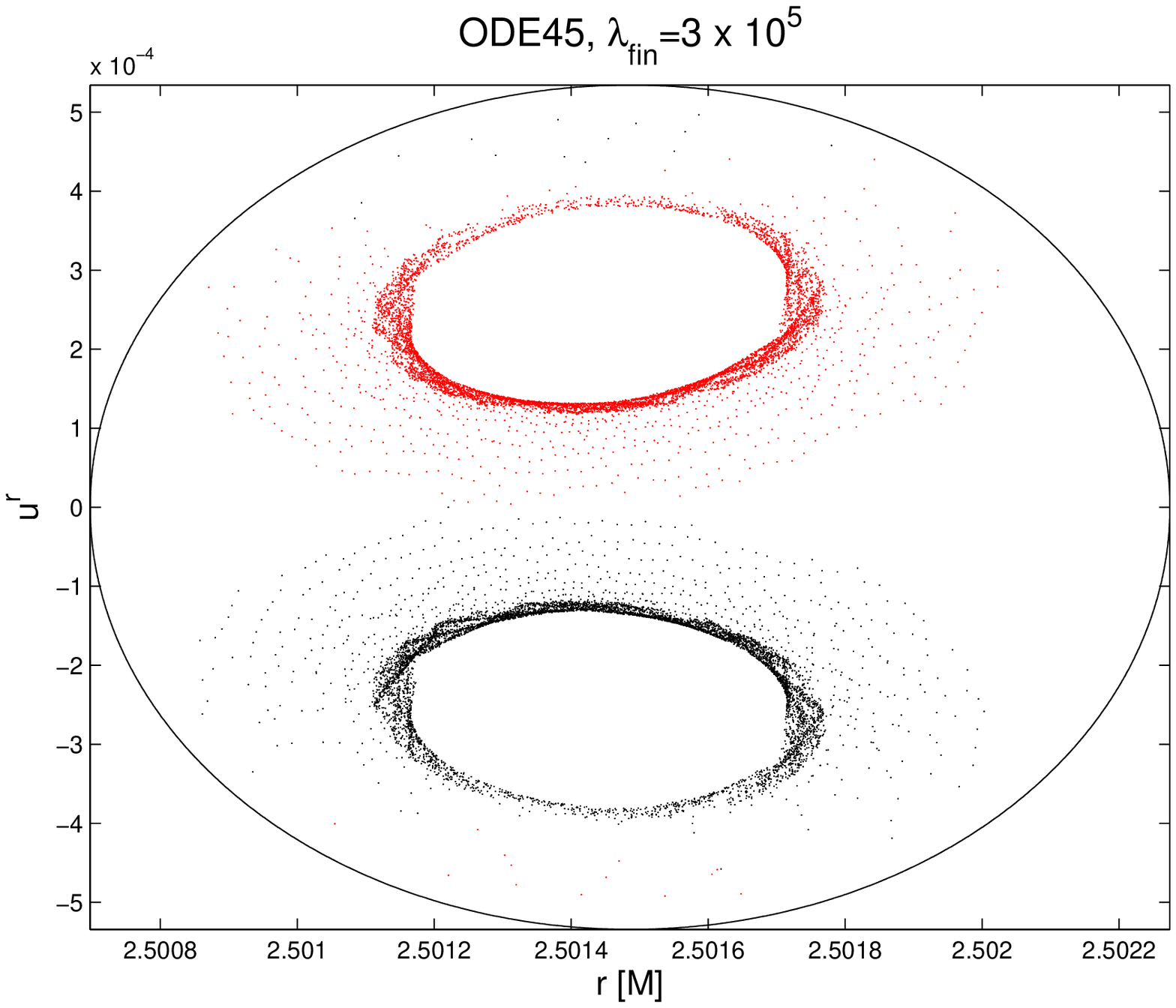}
\caption{We show how the accuracy of the integration crucially affects the appearance of the Poincar\'{e} surfaces of section of a single regular trajectory with $\tilde{q}\tilde{Q}=1.76$, $\tilde{L}=4.02\;M$ and $\tilde{E}=1.619855$ on the Kerr background
$a=0.55\;M$ with Wald magnetic field
$\tilde{q}B_{0}=1.92\:M^{-1}$. Particle is
launched at $r(0)=2.5012\:M$, $\theta(0)=1.0447$ with $u^r(0)=0$. We distinguish downward crossing with $u^{\theta}\geq0$ (black point) from the upward crossing with $u^{\theta}<0$ (red point) in the surfaces of section.}
\label{rezy}
\end{figure}

In the case of the chaotic trajectory (depicted in \rff{traj_chaos_3d}) the dynamics changes in favor of symplectic solver GLS. In \rff{traj_chaos} we observe that in this case the symplectic scheme is superior to the others in even more convincing manner than it was in the regular case. Although the initial phase when the error induced by GLS rises more steeply than that of ODE87 is also present, it turns over very quickly and for
$\lambda\gtrsim5\times10^3$ ($\approx 50$ azimuthal revolutions) the GLS turns out to be more accurate. 
The difference then rises much faster compared to the regular case. 

Experiments with ODE113 reveal that here we obtain distinct (though not sharply) errors by changing the RelTol. Difference of eight orders of magnitude in RelTol resulted in roughly one order difference in global error. We also note that chaotic regime induces disorder in short-time oscillations of the global error (see bottom panel of \rff{traj_chaos}). We summarize that the chaotic regime accents the supremacy of the symplectic scheme which is to be applied on all time scale here (except very short integrations where ODE87 dominates) to obtain the most accurate results. For fast though inaccurate calculation one would switch to ODE113 as before. Results for the chaotic orbit are summarized in Table \ref{tab_chaos}. 

From a practical point of view we demand high accuracy of the long-term integration when constructing Poincar\'{e} surfaces of section. By theory the intersection points with regular trajectory form one-dimensional curve in the section plane. In \rff{rezy} we observe, however, that the points may be dispersed over the considerable area if the global error in energy rises causing artificial excitation/dumping of the system. Symplectic integrator GLS provides the most reliable outcome, with ODE87 the curve is blurred significantly but the interpretation remains unambiguous. With ODE113 the curve is further blurred and using ODE45 solver we obtain completely unreliable outcome which could easily lead to the incorrect interpretation of a trajectory as a chaotic one. We note that we intentionally chose such trajectory which is highly sensitive to the relative errors in dynamic quantities since it itself spans small range of coordinate and momenta values.

\section{Conclusions}\label{conclus}
We confirm that the symplectic integrators are the method of choice in the case of long-term integration of the Hamiltonian system which in our case consists of a charged test particle orbiting around the Kerr black hole with stationary and axisymmetric electromagnetic test field. Its supremacy over non-symplectic methods is even more apparent in the case of chaotic orbits, where the global accuracy of non-symplectic methods decreases rapidly. The accuracy of the symplectic integrator could be further increased by reducing the stepsize (at the cost of the computational time). On the other hand the performance of the non-symplectic solvers is not considerably affected by reducing the local error (controlled by the RelTol parameter in our case) across the wide range of the values. Once the integrator does not fit the problem (= is not symplectic) there is no effective way to control the global error and even the extremely small local truncation errors do not ensure reliable outcome on a long time scale.

We suggest that our results are not problem-specific and may be generalized to the broad class of the systems. In particular, we suppose that symplectic integrators provide outstanding results in the chaotic regime of any non-integrable Hamiltonian system.

\ack
The present work was supported by the Grant Agency of Charles University (project GAUK 119210/2010) and Charles University research project SVV-263301. Authors also acknowledge the support from the Czech Science Foundation (project No.\ 205/09/H033).

\bibliography{\jobname}
\end{document}